\documentclass[fleqn,usenatbib,useAMS,twocolumn,a4paper,aps,nofootinbib]{revtex4}

\usepackage{aas_macros}
\usepackage{mathtools}
\usepackage{amsmath, amssymb}
\usepackage[T1]{fontenc}
\usepackage{ae,aecompl}
\usepackage{xcolor}
\usepackage{orcidlink}

\usepackage{graphicx}	
\usepackage{amsmath}	
\usepackage{amssymb}	
\usepackage{bm}		

\def\red{\color{red}}

\begin{document}

\title [Thermal conduction in magnetized disc]{Effect of thermal conduction on accretion shocks in relativistic magnetized flows around rotating black holes}

\author{Monu Singh}\email{monu18@iitg.ac.in}
\author{Camelia Jana}\email{camelia\_jana@iitg.ac.in}
\author{Santabrata Das}\email{sbdas@iitg.ac.in}
\affiliation{Department of Physics, Indian Institute of Technology Guwahati, Guwahati, 781039, Assam, India.}


\begin{abstract}
    We examine the effects of thermal conduction on relativistic, magnetized, viscous, advective accretion flows around rotating black holes considering bremsstrahlung and synchrotron cooling processes. Assuming the toroidal component of magnetic fields as the dominant one, we self-consistently solve the steady-state fluid equations to derive the global transonic accretion solutions for a black hole of spin $a_{\rm k}$. Depending on the model parameters, the magnetized accretion flow undergoes shock transitions and shock-induced global accretion solutions persist over a wide range of model parameters including the conduction parameter ($\Upsilon_{\rm s}$), plasma-$\beta$, and viscosity parameter ($\alpha_{\rm B}$). We find that the shock properties---such as shock radius ($r_{\rm s}$), compression ratio ($R$), and shock strength ($S$)---are regulated by $\Upsilon_{\rm s}$, plasma $\beta$, and $\alpha_{\rm B}$. Furthermore, we compute the critical conduction parameter ($\Upsilon_{\rm s}^{\rm cri}$), a threshold beyond which shock formation ceases to exist, and investigate its dependence on plasma-$\beta$ and $\alpha_{\rm B}$ for both weakly rotating ($a_{\rm k} \rightarrow 0$) and rapidly rotating ($a_{\rm k} \rightarrow 1$) black holes. Finally, we examine the spectral energy distribution (SED) of the accretion disc and observe that increased thermal conduction and magnetic field strength lead to more luminous emission spectra from black hole sources.
\end{abstract}


\maketitle

\section{Introduction}
\label{sec:intro}

Black holes are believed to be the central to some of the most energetic phenomena in the universe, such as active galactic nuclei (AGNs) and X-ray binaries (XRBs) \cite[and references therein]{remillard-etal-2006,netzer-2013}. The accretion process plays a key role in driving black hole dynamics, offering essential insights into the observational signatures of these enigmatic objects. Recent groundbreaking images of Sgr A$^*$ \cite[]{EHT-SgrA-2022} and M87 \cite[]{EHT-M87-2021} have revealed these supermassive black holes are surrounded by a hot, magnetized plasma. These AGNs are generally powered by advection-dominated accretion flows \cite[]{narayan-yi-1994,narayan-yi-1995,yuan-narayan-2014}, which exhibit a low-density profile that results in their characteristic low luminosity \cite[]{Ho-2008, netzer-2013}. This behavior is due to the inability of low-density plasma to cool efficiently through radiation, leading to flow temperatures that approach the virial temperature. Such high temperature and low density accretion flows are well-suited to explain the observational features of accreting super massive black hole systems \cite{yuan-narayan-2014}. Moreover, ubiquitous magnetic fields in accreting system plays pivotal role in describing the structure of the accretion flow as well \cite[]{Oda-etal-2007,sarkar-das-2016,das-sarkar-2018,jana-das-2024}.

In the hot, low density plasma of magnetized advective accretion flows, where the particle mean free path are significantly long, thermal conduction emerges as a crucial mechanism for energy transport. Hence, thermal conduction exerts a significant influence on the structure,and thermodynamic properties of hot accretion flows. Meanwhile, \cite{tanaka-menou-2006} estimated the mean free path using Chandra Observatory data from Sgr A* and proposed that accretion in this system occurs under weakly collisional conditions. In such regimes, the mean free path exceeds the characteristic length scale of the system, such as the gravitational radius ($r_{g}$). They also suggested that thermal conduction plays a pivotal role in driving the formation of outflows and winds. Furthermore, \cite[]{johnson-quataert-2007} and \cite[]{sharma-etal-2008} carried out comprehensive studies on the influence of electron thermal conduction in spherical hot accretion flows. Afterwards, numerous efforts were made using self-similar approach to emphasize the significant role of thermal conduction in shaping the behavior of black hole accretion systems in presence of bipolar outflows \cite[][]{tanaka-menou-2006,shadmehri-2008,Faghei2012,Khajenabi-Shadmehri-2013,ghoreyshi-shadmehri-2020}. Recently, \cite{mitra-etal-2023} studied the global accretion solutions and showed that the inclusion of thermal conduction notably modifies their transonic characteristics.

Accretion flow around black hole is inherently transonic as subsonic flow from the outer edge of the accretion disc eventually crosses the black hole horizon at supersonic speed. Such smooth transition happens at the critical points ($r_{\rm c}$). As the accretion progresses, rotating supersonic flow experiences centrifugal repulsion and flow is decelerated causing matter to accumulate in the vicinity of the black hole. When this accumulation exceeds a critical threshold, centrifugal barrier triggers a shock transition, provided Rankine-Hugoniot shock conditions are satisfied \cite[]{landau-lifshitz-1959}. The second law of thermodynamics further favors the formation of shock in a magnetized viscous advective accretion flows around black hole that naturally drives the accreting mater toward the state of higher entropy \cite{becker-kazanas-2001,mitra-das-2024}. The phenomenon of shock formation in accretion  around black holes has been extensively studied by various group of researchers, both theoretically and numerically considering hydrodynamics as well as magnetohydrodynamic scenarios \cite[]{Fukue1987,chakrabarti-1989, Chakrabarti-Molteni1993,molteni-etal-1994,yang-kafatos-1995,chakrabarti-1996,Molteni-etal1996,ryu-etal-1997,Lanzafame-etal1998,lu-etal-1999,das-etal-2001,becker-kazanas-2001,chakrabarti-das-2004,Le-Becker2004,Fukumura-Tsuruta2004,takahashi-etal-2006,das-2007,fukumura-etal-2007,becker-etal-2008,Das-etal2009,Kumar-etal2013,Das-etal2014,okuda-das-2015,sukova-janiuk-2015,fukumura-etal-2016,sarkar-das-2016,sukova-janiuk-2017,aktar-etal-2017,dihingia-etal-2018,dihingia-etal-2019A,kim-etal-2019,Okuda-etal2019,dihingia-etal-2020,Sen-etal2022,singh-das-2024,singh-das-2025,mitra-das-2024,Debnath-etal2024,TianLe-Zhao-etal2024,patra-etal-2024,singh-das-2024,jana-das-2024}. The shock-induced accretion solutions offer a compelling explanation for the spectro-temporal properties commonly observed in black hole candidates \cite[]{chakrabarti-titarchuk-1995,chakrabarti-manickam-2000,mandal-chakrabarti-2005,nandi-etal-2012,iyer-etal-2015,das-etal-2021,majumder-etal-2022,nandi-etal-2024}. However, the role of thermal conduction in shock-induced, magnetized, viscous, advective accretion flows in presence of radiative cooling processes remains unexplored in the astrophysical literature.

Inspired by this, in the present study, we investigate the structure of a steady, magnetized, viscous, advective accretion flows around rotating black holes incorporating the effects of thermal conduction along with bremsstrahlung and synchrotron cooling mechanisms. We consider that accretion flow is threaded by toroidal magnetic fields surrounding a Kerr black hole, with the gravitational field approximated using the pseudo-potential formulation \cite[]{dihingia-etal-2018}. Moreover, we employ a relativistic equation of state (REoS) that effectively accounts for the thermodynamic variables of the accretion flow. Using this framework, we derive the shock-induced global accretion solutions and investigate key shock properties, such as shock location ($r_{\rm s}$), compression ratio ($R$), and shock strength ($S$), while examining their dependence on thermal conduction and magnetic fields. Finally, we analyze the influence of model parameters on the  emission spectrum of supermassive black holes.

The plan of this paper is as follows: In Section II, we outline the basic assumptions, governing equations and solution methodology. Obtained results are presented in Section III. Finally, in Section IV, we summarize our results.

\section{Governing equations and methodology}

We begin with a magnetized, viscous, advective, and axisymmetric low angular momentum accretion flows around a rotating black hole in the presence of thermal conduction. We adopt effective pseudo-potential \cite[]{dihingia-etal-2018} to mimic the space-time geometry around the rotating black hole. This approach allows us to avoid the complexities of general relativistic prescription, yet retaining the effects of background space-time geometry around black hole, and maintaining the accretion energetics \cite{dihingia-etal-2018,das-etal-2021,singh-das-2024,jana-das-2024,kumar-etal-2025}. We utilize a cylindrical coordinate system ($r,\phi,z$) and assume that the flow maintains hydrostatic equilibrium in the vertical ($z$) direction. We choose $G = M_{\rm{BH}} = c = 1$ to express the flow variables in dimensionless form, where $G$ is the gravitational constant, $M_{\rm{BH}}$ is mass of the black holes, and $c$ is the speed of light. In this unit system, length, angular momentum, and energy are expressed in units of $G M_{\rm{BH}}/c^2$, $G M_{\rm{BH}}/c$, and $c^2$, respectively. With this, we write the governing equations that describe the motion of the accretion flow around rotating black holes as \cite{dihingia-etal-2018}:

\begin{align}    
    &\upsilon \frac{d\upsilon}{dr}+\frac{1}{h \rho}\frac{dP}{dr}+\frac{d \Psi_{\textrm{e}}^{\textrm{eff}}}{dr} + \frac{\langle B_{\phi}^2 \rangle}{4 \pi r \rho}  = 0,\label{eq:rad-mom}\\
    &\dot{M} = 2\pi \upsilon \Sigma \sqrt{\Delta},\label{eq:acc-rate}\\
    &\upsilon \frac{d\lambda}{dr} + \frac{1}{\Sigma\space r}\frac{d}{dr}(r^2 T_{r\phi}) = 0,\label{eq:azi-mom}\\    
    &\frac{\upsilon}{\rho(\Gamma - 1)}\left(\frac{dP_{\rm gas}}{dr} - \frac{\Gamma P_{\rm gas}}{\rho}\frac{d\rho}{dr}\right) = \Lambda - \frac{1}{r\rho}\frac{d(rF_{\rm s})}{dr} \nonumber\\ 
    & \qquad \qquad \qquad \qquad \qquad \qquad -\alpha_{\rm B} r (P/\rho+ \upsilon^2)\frac{d\Omega}{dr} ,\label{eq:entroy}\\
    &\frac{\partial \langle B_{\phi} \rangle \hat{\phi}}{\partial t} = \nabla \times \left( \vec{\upsilon} \times \langle B_{\phi} \rangle \hat{\phi} - \frac{4\pi}{c} \eta \vec{j} \right),\label{eq:induction}
\end{align}
where $\upsilon$ is the radial velocity, $\rho$ is the mass density, $h~[= (\epsilon + P_{\rm gas})/\rho]$ is the enthalpy,  $B_{\phi}$ denotes the azimuthal component of magnetic fields, and `$\langle ~ \rangle$' represents the azimuthal average. The total isotropic pressure $P$ is the sum of the gas pressure ($P_{\rm gas}$) and magnetic pressure ($P_{\rm mag}$) with $P_{\rm gas} = R \rho T/\mu$, $R$, $T$ and $\mu$ being the universal gas constant, local flow temperature and mean molecular weight, respectively. In this work, we choose $\mu=0.5$ for fully ionized plasma. Further, we calculate $P_{\rm{mag}} = \frac{\langle B_{\phi}^2 \rangle}{8 \pi}$ and rewrite  the total pressure as $P = P_{\rm{gas}}(1+1/\beta)$ where the plasma-$\beta$ parameter is defined as  $\beta= P_{\rm{gas}}/P_{\rm{mag}}$. In equation (\ref{eq:rad-mom}), $\Psi_{\rm e}^{\rm eff}$ is the effective potential on the  equatorial plane due to a rotating black hole, which is given by \cite{dihingia-etal-2018},
\begin{equation}
    \Psi_{\rm e} ^{\rm eff} = 1+\frac{1}{2}\ln\left[\frac{r \Delta}{a_{\rm k}^2 (r+2) - 4 a_{\rm k} \lambda + r^3 -\lambda^2(r-2)}\right],
\end{equation}
where $\lambda$ and $a_{\rm k}$ denote the specific angular momentum of the flow and spin of the black hole, respectively, and $\Delta = r^2 - 2 r + a_{\rm k}^2$. 

In equation (\ref{eq:acc-rate}), $\dot M$ denotes the mass accretion rate, which remains constant throughout the flow. We express mass accretion rate as ${\dot m} = {\dot M}/{\dot M}_{\rm Edd}$, where ${\dot M}_{\rm Edd}~(=1.44\times 10^{17} (M_{\rm BH}/M_\odot)$ gm s$^{-1})$, $M_\odot$ being solar mass. Here, $\Sigma$ refers to the vertically integrated surface mass density of the accreting matter, given by  $\Sigma = 2 \rho H$ \cite{matsumoto-1984}, where $H$ is the local half-thickness of the disc. We calculate $H$ as described in \cite{Riffert-Herold-1995, peitz-Appl-1997} and is given by,
\begin{equation}
  H = \sqrt{ \frac{P r^3 (1-\lambda \Omega) }{\rho} \times  \frac{(r^2 + a_{\rm k}^2)^2 - 2 \Delta a_{\rm k}^2}{(r^2 + a_{\rm k}^2)^2 + 2 \Delta a_{\rm k}^2}}, 
\end{equation}
where $\Omega$ is angular velocity of the flow and is given by,
$$
\Omega = \frac{2 a_{\rm k} + \lambda (r - 2)}{a_{\rm k}^2(r + 2) - 2 a_{\rm k}\lambda + r^3}.
$$

In equation (\ref{eq:azi-mom}), we consider the vertically integrated total stress, which is predominantly dominated by the $r\phi$ component of the Maxwell stress $T_{r \phi}$, outweighing the contributions from the other components. For an advective flow with substantial radial velocity, we calculate $T_{r \phi}$ as \cite{Machida-etal-2006}, $ T_{r\phi}= \frac{\langle B_{r} B_{\phi} \rangle}{4 \pi} H =  - \alpha_{\rm B} (W + \Sigma \upsilon^2)$, where $W$ is the vertically integrated pressure and $\alpha_{\rm B}$ is a constant that regulates the viscous effect inside the disc.

In this work, we assume that the flow is cooled through both bremsstrahlung and synchrotron processes. The bremsstrahlung and synchrotron rates (in units of erg cm$^{-3}$ s$^{-1}$) are given by \cite{shapiro-teukolsky-1983, rybicki-lightman-1986, wardzinski-zdziarski-2001},
\begin{align*}
    & Q^{\rm brem} = 1.4 \times 10^{-27} n_{e}^2 T_{e}^{1/2}(1+4.4\times 10^{-10} T_{e}),\\
    & Q^{\rm syn} = \frac{16}{3}\frac{n_{e} e^2}{c}\left(\frac{e B}{m_{e} c}\right)^2\left(\frac{k_{B} T_{e}}{m_{e} c^2}\right)^2,
\end{align*}
where, $n_{e}$ is electron number density, $T_{e}$ is the electron temperature, $e$ is the electron charge, $m_e$ is the electron mass, $k_B$ is the Boltzmann constant and $B$ is the magnetic fields. With this, we obtain the dimensionless total cooling rate as $\Lambda=(Q^{\rm brem} + Q^{\rm syn})/\rho ~ \times ~(GM_{\rm BH}/c^5)$. Following \cite{cowie-mcKee-1977,tanaka-menou-2006}, we estimate the saturated conduction flux as $F_{\rm s} = 5 \Upsilon_{\rm s} \rho \left(\frac{P_{\rm gas}}{\rho}\right)^{3/2}$, where $\Upsilon_{\rm s}$ is the dimensionless saturated conduction parameter (hereafter conduction parameter) lies in the range $0 \le \Upsilon_{\rm s} < 1 $. Meanwhile, earlier studies showed that self-similar accretion solutions tend to become non-rotating ($\Omega \rightarrow 0$) as the conduction parameter $\Upsilon_{\rm s}$ approaches its limiting value \cite[and references therein]{shadmehri-2008,ghasemnezhad-2018,mitra-etal-2023}. This limiting value typically lies well below unity, ensuring physically acceptable accretion solutions around a black hole \cite[]{mitra-etal-2023,singh-das-2025}. Here, $\Gamma$ is the adiabatic index of the accreting gas.

The advection rate of the toroidal magnetic field is described using the induction equation, with its azimuthally averaged form presented in equation \ref{eq:induction}. Here, $\vec{v}$ denotes the velocity vector, $\eta$ represents the resistivity, and $j$ is current density, respectively \cite{Oda-etal-2007}. Due to the large scale of the accretion , the Reynolds number is typically high, allowing us to neglect the magnetic diffusion term. Furthermore, we disregard the dynamo term and assume that the azimuthally averaged toroidal magnetic fields approach zero at the  surface. As a result, the toroidal magnetic flux rate can be expressed as:
\begin{equation}
    \dot{\Phi} = -4\pi \upsilon H B_{0}(r),
\end{equation}
where, $B_{0}(r)$ is the azimuthally averaged toroidal magnetic field confined to the disc's equatorial plane. The magnetic flux $\dot{\Phi}$ is not conserved in the accretion flow and may vary inversely with the radial distance $r$. Following \cite{Machida-etal-2006,Oda-etal-2007}, we consider $\dot{\Phi} \propto r^{-\zeta}$, where $\zeta$ is a parameter representing the magnetic flux advection rate. Taking all these factors into account, the parametric relation for the magnetic flux rate is:
$$
\dot{\Phi}(r,\zeta,\dot{M}) = \dot{\Phi}_{\rm{edge}}(\dot{M})\left( \frac{r}{r_{\rm{edge}}} \right)^{-\zeta},
$$
where $\Phi_{\rm{edge}}$ is the advection rate of toroidal magnetic flux obtained at the outer edge of the disc ($r_{\rm{edge}}$). In this work, we consider $\zeta=1$ unless stated otherwise. 

The flow equations are closed using the relativistic equation of state (REoS) that relates the internal energy $(\epsilon)$, $P_{\rm{gas}}$ and $\rho$ of the accretion flow. To this end, we adopt the REoS proposed by \cite{chattopadhyay-ryu-2009}, which is given by,
\begin{equation}
\epsilon = \frac{\rho f}{\left(1+\frac{m_p}{m_e}\right)},
\end{equation}
with
\begin{equation}
f = \left[1 + \Theta \left(\frac{9\Theta + 3}{3\Theta+2}\right) \right] + \left[\frac{m_p}{m_e} + \Theta \left(\frac{9\Theta m_e + 3 m_p}{3\Theta m_e + 2 m_p}\right) \right],
\end{equation}
where $m_{\rm p}$ is the mass of the ions and $\Theta (= k_B T/m_e c^2)$ is the dimensionless temperature. Using the REoS, we have the polytropic index as $N =(1/2)(df/d\Theta)$ and adiabatic index $\Gamma = (1+N)/N$. Subsequently, we define the sound speed as $C_{\rm s} = \sqrt{ \Gamma P_{\rm gas}/(\epsilon + P_{\rm gas}) }  = \sqrt{ 2\Gamma \, \Theta/(f + 2\Theta)}$.

We simplify equations (\ref{eq:rad-mom})- (\ref{eq:induction}) to obtian the wind equation along with the radial gradients of $\Theta$, $\lambda$ and $\beta$ as
\begin{align}
    & \frac{d\upsilon}{d r} = \frac{\mathcal{N}(r,\upsilon,\lambda,\Theta)}{\mathcal{D}(r,\upsilon,\lambda,\Theta)}, \label{dvdr}\\
    & \frac{d\Theta}{d r} = \Theta_{\textrm{1}} + \Theta_{\textrm{2}}\frac{d\upsilon}{d r}, \label{dthetadr}\\
    & \frac{d\lambda}{d r} = \lambda_{\textrm{1}} + \lambda_{\textrm{2}}\frac{d\upsilon}{d r},\label{dlambdadr}\\
    & \frac{d\beta}{d r} = \beta_{\textrm{1}} + \beta_{\textrm{2}}\frac{d\upsilon}{d r}.\label{dbetadr}
\end{align}
In equations (\ref{dvdr}, \ref{dthetadr}, \ref{dlambdadr}, \ref{dbetadr}), the explicit expressions for the quantities $\mathcal{N}$, $\mathcal{D}$, $\Theta_{1}$, $\Theta_{2}$, $\lambda_{1}$, $\lambda_{2}$, $\beta_{1}$, and $\beta_{2}$ are mathematically extensive. Therefore, we provide them in the Appendix.

In order to obtain the global accretion solutions, we simultaneously solve equations (\ref{dvdr}, \ref{dthetadr}, \ref{dlambdadr}, \ref{dbetadr}) \cite[and references therein]{das-2007,sarkar-das-2016,das-sarkar-2018,singh-das-2024}. In doing so, we consider mass accretion rate ($\dot{m}$), viscosity parameter ($\alpha_{\rm{B}}$), black hole spin ($a_{\rm k}$), and conduction parameter ($\Upsilon_{\rm s}$) as global parameters, since these quantities remain constant throughout. During accretion, the subsonic flow begins its journey from the outer edge of the disc ($r_{\rm edge}$) and gradually moves toward the black hole. As it progresses, the flow reaches the critical point ($r_{\rm c}$), where it makes smooth transition to supersonic state before entering the black hole. Due to the inherently transonic nature of black hole accretion, the flow must pass through the critical point ($r_{\rm c}$). Hence, we choose $r_{\rm c}$ as reference radius ($r_{\rm ref}$) and supply angular momentum ($\lambda_c$) and plasma-$\beta$ ($\beta_c$) at $r_{\rm c}$ as local flow parameters. Employing the model parameters, we carry out the critical point analysis and apply l$'$H\^{o}pital's rule to calculate the radial velocity gradient, which takes the form $(dv/dr)_{\rm c}=0/0$ at $r_{\rm c}$. We then solve $\mathcal{N}(r,\upsilon,\lambda,\Theta)=0$ and $\mathcal{D}(r,\upsilon,\lambda,\Theta)=0$ to find radial velocity ($\upsilon_{c}$) and temperature ($\Theta_{c}$) at $r_{\rm c}$. Subsequently, using these flow variables at $r_{\rm c}$, we integrate equations (\ref{dvdr}), (\ref{dthetadr}), (\ref{dlambdadr}), (\ref{dbetadr}) up to horizon ($r_{\rm h}$) and then to the outer edge of the disc ($r_{\rm edge}$). Finally, we join these two parts of the solution to obtain the global accretion solution, noting the flow variables at $r_{\rm edge}$ as energy $\mathcal{E}_{\rm edge}$, angular momentum $\lambda_{\rm edge}$, plasma-$\beta_{\rm edge}$, radial velocity $\upsilon_{\rm edge}$, and temperature $\Theta_{\rm edge}$. It is important to note that the same global accretion solution can also be derived using the outer boundary flow variables. In the following sections, we will continue to present our results based on these outer boundary flow variables.

Depending on the model parameters, the flow may exhibit either single or multiple critical points. Critical points located near the horizon ($r_{\rm h}$) are referred to as inner critical points ($r_{\rm{in}}$), while those formed at a large distance are known as outer critical points ($r_{\rm{out}}$). Notably, accretion flows with multiple critical points are of particular interest, as they can undergo shock transitions. Based on the extreme physical conditions, three types of shocks are possible, namely Rankine-Hugoniot shocks, isentropic shocks and isothermal shocks \cite[]{Abramowicz-Chakrabarti1990}. In this work, we study Rankine-Hugoniot, assuming that no energy is lost through the flow surface at the shock radius and it occurs provided Rankine-Hugoniot shock conditions (RHCs)\footnote{RHCs are the conservation of (a) mass flux ${\dot M}_{+}={\dot M}_{-}$, (b) energy flux ${\mathcal{E}}_{+}={\mathcal{E}}_{-}$, (c) momentum flux $W_{+}+\Sigma_{+}v^2_{+}=W_{+}+\Sigma_{+}v^2_{+}$ and (d) magnetic flux ${\dot \Phi}_{+}={\dot \Phi}_{-}$ across the shock front \cite[]{landau-lifshitz-1959,sarkar-das-2016,das-sarkar-2018,singh-das-2025,jana-das-2024}. Here, `$-/+$' refer upstream/downstream quantities across the shock front, while $\mathcal{E}~[=v^2/2+\log h+\Psi_{\rm e}^{\rm eff}+ \langle B_{\phi}^2 \rangle/(4 \pi \rho)]$ denotes the local flow energy.} are satisfied. In reality, rotating accreting matter experiences centrifugal repulsion in the vicinity of the black hole, resulting in an accumulation of matter that forms an effective boundary layer around it. However, this accumulation cannot continue indefinitely, as it leads to the continuous transitions in the flow variables, manifesting as shock waves at its threshold. This shock transition is in accordance with the second law of thermodynamics, as it is characterized by a higher entropy state of the accreting matter \cite[]{becker-kazanas-2001}. Due to shock compression, the density and temperature of the convergent flow sharply increase downstream, just after the shock transition, resulting in a hot and dense post-shock flow (hereafter referred to as the post-shock corona, PSC). As a result, the PSC serves as an ideal location for reprocessing soft photons from the pre-shock flow into high energy X-ray radiation through the inverse Comptonization process. These high energy X-rays are commonly observed from black hole X-ray binary (BH-XRB) sources \cite[and references therein]{chakrabarti-titarchuk-1995,mandal-chakrabarti-2005,nandi-etal-2012,iyer-etal-2015,Nandi-etal2018,majumder-etal-2022}.

\section{Result}

\begin{figure}
\begin{center}
    \includegraphics[width=\columnwidth]{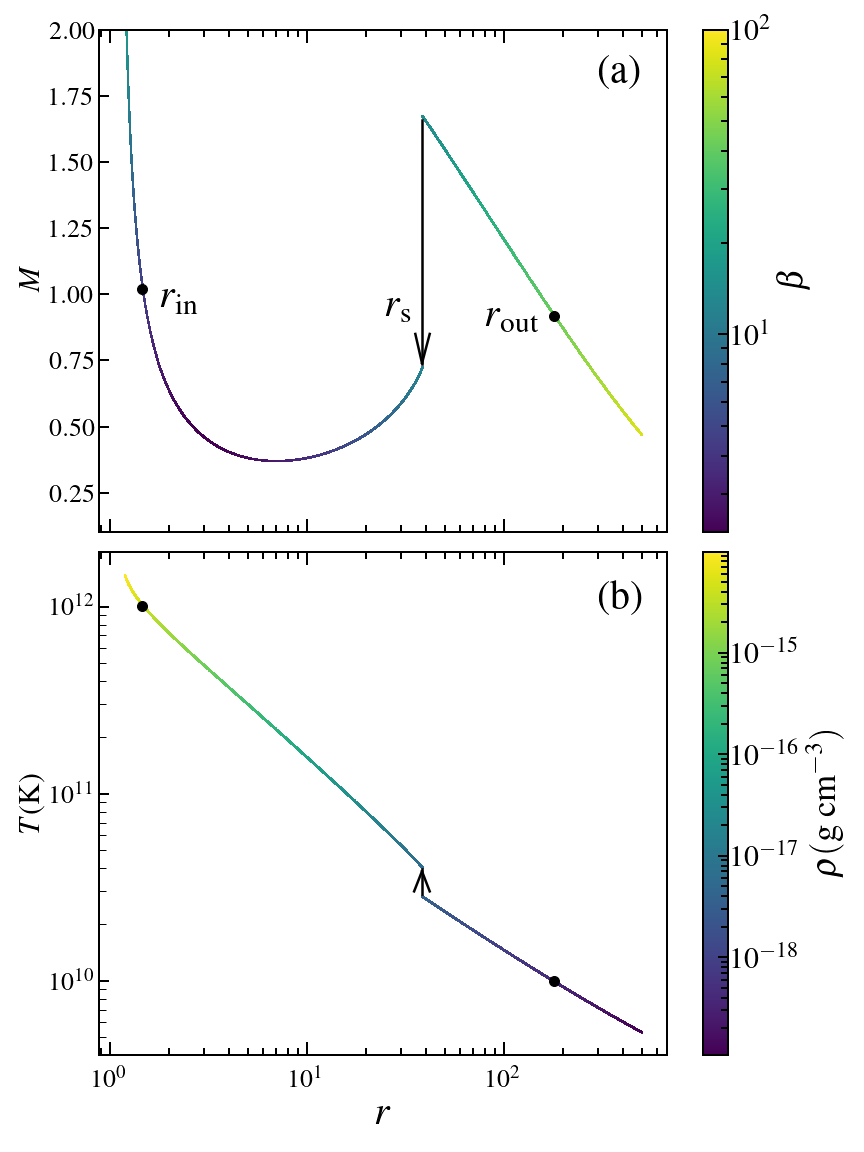}
\end{center}
\caption{Example of a global shocked-induced accretion solution, where in panel (a) we plot the variation of Mach number ($M=\upsilon/C_{\rm{s}}$) and plasma-$\beta$ (in color) with radial coordinate ($r$), and panel (b) illustrates the temperature ($T$) of the flow along with the density ($\rho$, in color). Here, flow is injected from the outer edge of the  $r_{\rm{edge}}=500$ with $\lambda_{\rm{edge}}=2.266$, $\mathcal{E}_{\rm{edge}} = 1.00105$, $\beta_{\rm{edge}}=80$, $\alpha_{\rm{B}} = 0.01$, $\dot{m}=0.0001$, $\Upsilon_{\rm s}=0.01$, and $a_{\rm{k}} =0.99$. Flow experiences shock transition at $r_{\rm s}=38.37$ indicated by the vertical arrow. Filled circles denote the critical points ($r_{\rm in}$ and $r_{\rm out}$). Color bars in panel (a) and (b) refer the ranges of $\beta$ and $\rho$. See the text for details.
}
\label{fig:1}
\end{figure}

In Fig. \ref{fig:1}, we depict an example of a shock-induced global magnetized accretion solution around a rotating black hole in presence of thermal conduction. In panel (a), we show the variation of Mach number ($M=\upsilon/C_{\rm{s}}$) with  radius ($r$). Here, we choose $\dot{m}=0.0001$, $\alpha_{\rm{B}} = 0.01$ and $\Upsilon_{\rm s}=0.01$ as global parameters, and inject flow subsonically from the outer edge of the disc ($r_{\rm{edge}}$) on to a super massive black hole of mass $M_{\rm{BH}} = 4.3 \times 10^{6} \: \mathrm{M}_{\odot}$ and spin $a_{\rm{k}} =0.99$, with local flow parameters, namely angular momentum $\lambda_{\rm{edge}}=2.266$, energy $\mathcal{E}_{\rm{edge}} = 1.00105$ and plasma-$\beta_{\rm{edge}}=80$ at $r_{\rm{edge}}=500$. The flow becomes supersonic after passing the outer critical point at $r_{\rm{out}} = 180.28$ with angular momentum $\lambda_{\rm{out}}=2.056$, and continues to accrete toward the horizon. Meanwhile, RHCs become favorable, and the supersonic upstream flow undergoes a shock transition to the subsonic branch at $r_{\rm{s}}=38.37$, indicated by the vertical arrow. In this work, we consider the shock to be thin and non-dissipative \cite[]{Frank-etal2002}. After the shock, the temperature of the downstream flow is increased as the kinetic energy of the upstream flow is converted into thermal energy in the downstream. Furthermore, due to shock compression, the convergent flow becomes compressed, resulting in an increase in density in the downstream flow (PSC). In this work, we treat $\Upsilon_{\rm s}$ as a global parameter for simplicity, although it may be different depending on the degree of thermal conduction in the upstream and downstream flows. We refrain considering different $\Upsilon_{\rm s}$ to avoid introducing additional model parameters. In Fig. \ref{fig:1}b, we show the temperature ($T$) profile of the global shocked accretion solution presented in Fig \ref{fig:1}a, with density ($\rho$) variation indicated by colors. The range of the flow density is displayed in the color bar on the right side of panel (b). The density compression across the shock front is characterized by the compression ratio, defined as $R=\Sigma_{+}/\Sigma_{-}$, whereas the temperature jump is quantified by the shock strength, which is defined as $S=M_{-}/M_{+}$. For the shocked solution presented in Fig. \ref{fig:1}, we obtain $R=1.93$ and $S=2.31$.

\begin{figure}
\begin{center}
    \includegraphics[width=\columnwidth]{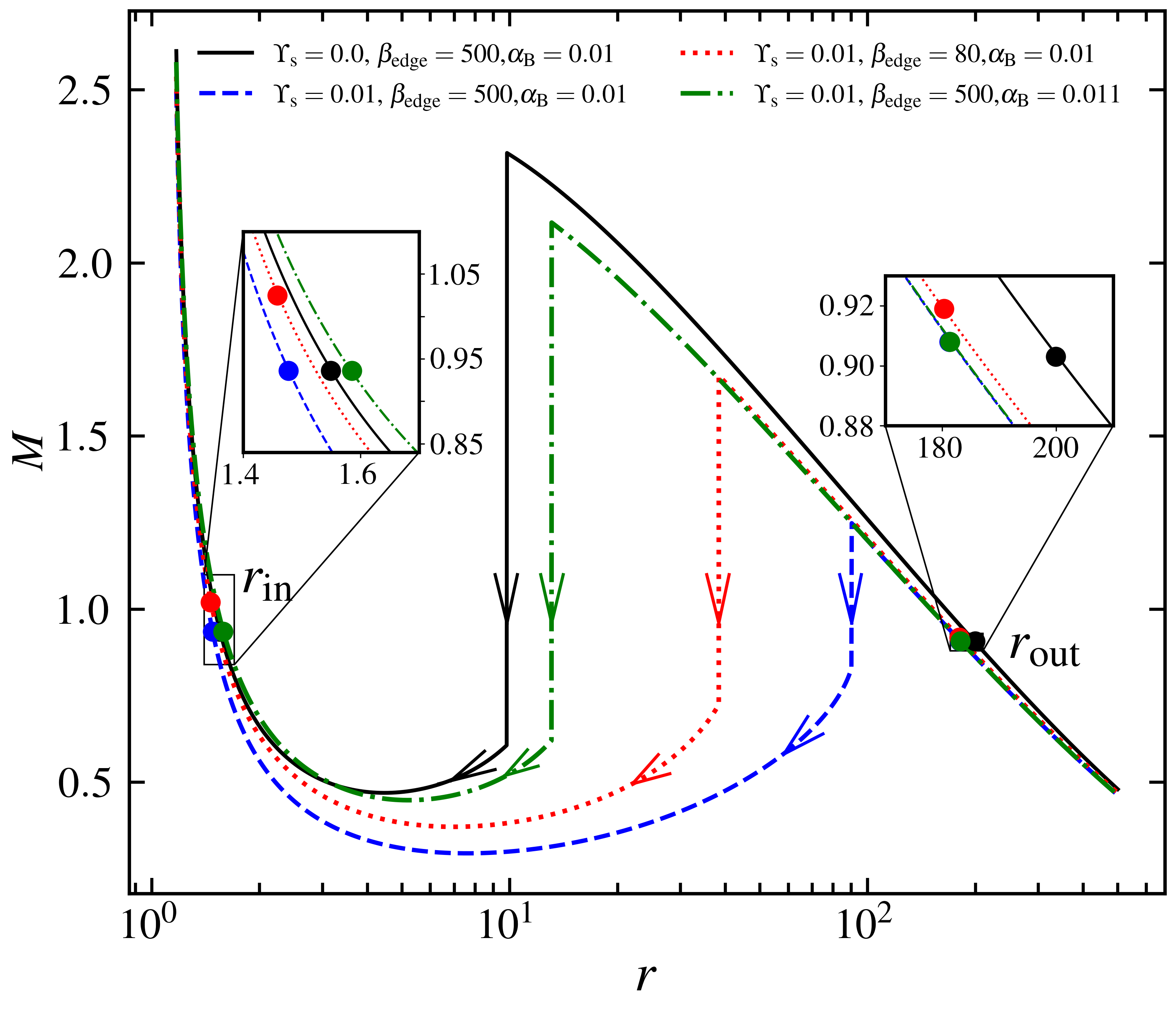}
\end{center}
\caption{Plot of Mach number $M=\upsilon/C_{\rm s}$ with the radial distance $r$. Flows of fixed $\mathcal{E}_{\rm edge} = 1.00105$, $\lambda_{\rm edge}=2.266$ are injected from $r_{\rm{edge}}=500$ with different $\Upsilon_{\rm s}$, $\beta_{\rm edge}$ and $\alpha_{\rm B}$ values. Here, $a_{\rm{k}} =0.99$ and $\dot{m}=0.0001$. Vertical arrows indicate the corresponding shock radii at $r_{\rm s}=9.82$ (solid), $90.24$ (dashed), $38.37$ (dotted) and $13.09$ (dot-dashed). Inner ($r_{\rm in}$) and outer ($r_{\rm out}$) critical points are zoomed at the insets for clarity. See the text for details.
}
\label{fig:2} 
\end{figure}

We now investigate the combined influence of viscosity, thermal conduction, and magnetic fields on the shock transition in accretion flows with a fixed outer boundary. In this analysis, we inject matter towards the black hole from the outer edge at $r_{\rm edge}=500$ with a local energy $\mathcal{E}_{\rm edge} = 1.00105$ and angular momentum $\lambda_{\rm edge}=2.266$. The mass accretion rate is set to ${\dot m}= 10^{-4}$, and the Kerr parameter is chosen as $a_{\rm k}=0.99$. For the set of parameters ($\alpha_{\rm B},~\beta_{\rm edge}, ~\Upsilon_{\rm s})~=~(0.01,~500,~0.0)$, a shock is formed at $r_{\rm s}=9.82$, as indicated by the vertical arrow and the shocked accretion flow solution is shown by the solid (black) curve in Fig. \ref{fig:2}. For this solution, we obtain the compression ratio $R=2.74$ and $S=3.81$. The flow passes through both inner and outer critical points at $ r_{\rm in}=1.549$ and $r_{\rm out}=199.91$ which are marked using filled circles. When thermal conduction is introduced ($\Upsilon_{\rm s}=0.01$) while keeping the other model parameters fixed, the shock front moves outward to $r_{\rm s}=90.24$. This happens because thermal conduction enhances the local thermal pressure, which in turn pushes the shock to settle at a larger radius \cite{singh-das-2025}. For this case, we find the inner and outer critical points at $r_{\rm in}=1.477$, and $r_{\rm out}=188.25$, and $(R,~S)~=~ (1.39,1.50)$, with the resulting solution represented by the dashed (blue) curve. As the magnetic field strength is increased to $\beta_{\rm edge}=80$ with $\Upsilon_{\rm s}=0.01$ and $\alpha_{\rm B}=0.01$, we observe that the shock front moves toward the horizon as indicated by the dotted red vertical arrow at $r_{\rm s}=38.37$. Here, the inner and outer critical points are at $r_{\rm in}=1.458$, and $r_{\rm out}=180.28$, and $(R,~S)~=~ (1.93,2.31)$. This inward movement of the shock is expected, as the density and temperature in the post-shock region (PSC) are higher than in the upstream flow. This results in more intense cooling, which reduces the thermal pressure, ultimately causing the shock to move inward. Overall, it is evident that thermal conduction induces effect opposite to the magnetic fields in determining the shock transitions. Finally, when viscosity is increased ($\alpha_{\rm B}=0.011$), while maintaining $\Upsilon_{\rm s}=0.01$ and $\beta_{\rm edge}=500$, we observe that the shock moves further inward to $r_{\rm s}=13.09$, as indicated by the dot-dashed (green) vertical arrow. The increased viscosity facilitates more efficient angular momentum transport, weakening the centrifugal repulsion and causing the shock to move inward. For this solution, we find the inner and outer critical points at $r_{\rm in}=1.582$, and $r_{\rm out}=180.36$, and $(R,~S)~=~ (2.55, 3.40)$. We tabulate the model parameters and shock properties in Table \ref{tab:table-1}. With these findings, we point out that the combined effects of thermal conduction, viscosity, and magnetic fields regulate the shock properties, including the size of the post-shock corona ($r_{\rm s}$), density compression ($R$) and temperature jump ($S$). 

\begin{table*}
\centering
    \caption{Model parameters, flow variables and shock properties for shock-induced global accretion solutions presented in Fig. \ref{fig:2} are tabulated. In columns $1-9$, conduction parameter ($\Upsilon_{\rm s}$), plasma-$\beta_{\rm edge}$, viscosity parameter ($\alpha_{\rm B}$), inner critical point ($r_{\rm in}$), angular momentum ($\lambda_{\rm in}$) at $r_{\rm in}$, plasma-$\beta_{\rm in}$, outer critical point ($r_{\rm out}$), angular momentum ($\lambda_{\rm out}$) at $r_{\rm out}$, plasma-$\beta_{\rm out}$, shock radius ($r_{\rm s}$), compression ratio ($R$), and shock strength ($S$) are presented. See text for the details.}
    \begin{tabular}{l c c c c c c c c c c c} 
    \hline \hline
    $\Upsilon_{\rm s}$ & $\beta_{\rm{edge}}$ & $\alpha_{\rm{B}}$ & $r_{\textrm{in}}$ &            $\lambda_{\textrm{in}}$ & $\beta_{\rm{in}}$ & $r_{\textrm{out}}$ & $\lambda_{\textrm{out}}$ & $\beta_{\rm{out}}$ & $r_{\textrm{s}}$ & $R$ & $S$ \\
    & & & ($r_{\textrm{g}}$) & ($r_{\textrm{g}} c$)& & ($r_{\textrm{g}}$)& ($r_{\textrm{g}} c$)& &($r_{\textrm{g}}$) \\ \hline
    0 & 500 & 0.01 & 1.549 & 1.944 & 20.195 & 199.91 & 2.070 & 274.711 & 9.82 & 2.74 & 3.81 \\
    0.01 & 500 & 0.01 & 1.477 & 1.947 & 27.328 & 181.25 & 2.054 & 266.812 & 90.24 & 1.39 & 1.50 \\ 0.01 & 80 & 0.01 & 1.458 & 1.949 & 4.531 & 180.28 & 2.056 & 42.512 & 38.37 & 1.93 & 2.31 \\
    0.01 & 500 & 0.011 & 1.582 & 1.915 & 25.770 & 181.36 & 2.039 & 267.013 & 13.09 & 2.55 & 3.40 \\
    \hline \\
    \end{tabular}
    \label{tab:table-1}
\end{table*}

\begin{figure}
\begin{center}
    \includegraphics[width=\columnwidth]{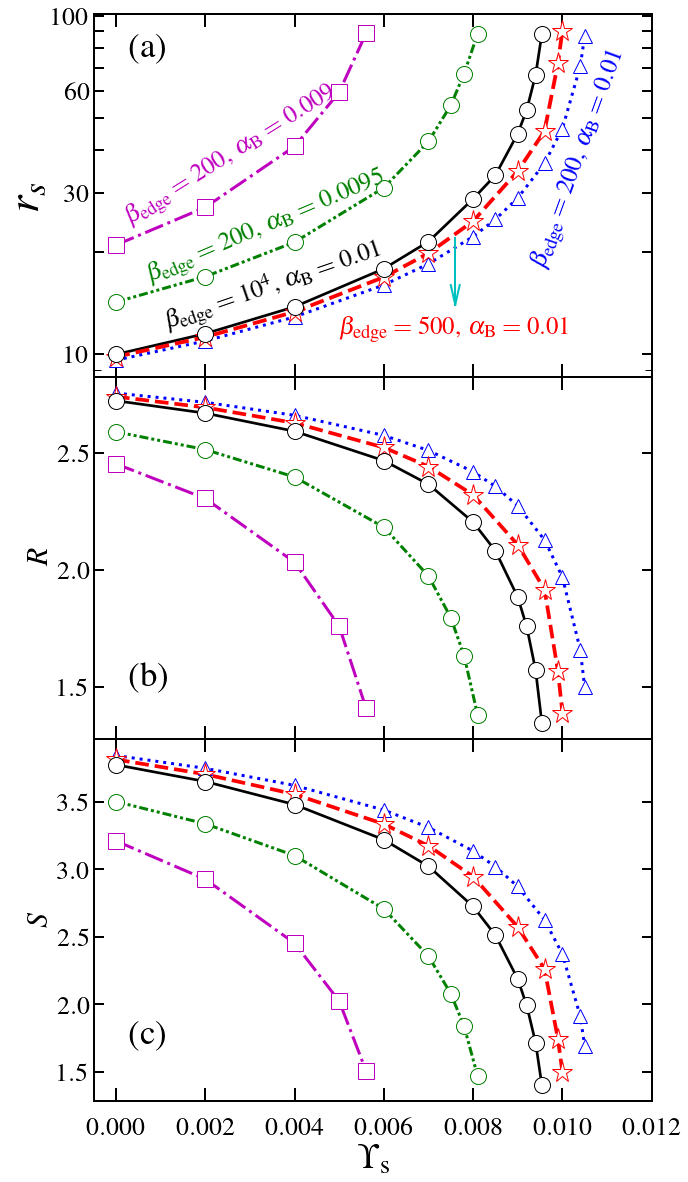}
\end{center}
\caption{Variations of the shock properties with $\Upsilon_{\rm s}$ for various values of the $\beta_{\rm edge}$ and $\alpha_{\rm B}$. In panel (a), we plot the shock radius ($r_{\rm s}$); in panel (b), the compression ratio ($R$), and in panel (c), we illustrate the shock strength ($S$). We choose $\alpha_{\rm B} = 0.01$, and $a_{\rm{k}} =0.99$. See the text for details.}
\label{fig:3} 
\end{figure}

Next, we compare the size of the post-shock corona (PSC) by examining the shock radius ($r_{\rm s}$) as a function of conduction parameter ($\Upsilon_{\rm s}$) for flows with varying viscosity ($\alpha_{\rm B}$) and magnetic field strengths ($\beta_{\rm edge}$). Here, matter is injected from $r_{\rm edge}=500$ with $\mathcal{E}_{\rm edge}=1.00105$ and $\lambda_{\rm edge}=2.266$ onto a black hole with spin $a_{\rm k}=0.99$. Initially, we fix $\beta_{\rm edge}=200$ at $r_{\rm edge}=500$ and vary $\alpha_{\rm B}$. The obtained results are presented in Fig. \ref{fig:3}(a), where open squares, circles and triangles connected with dot-dashed, dot-dot-dashed, and dotted lines correspond to $\alpha_{\rm B}=0.009$, $0.0095$ and $0.01$, respectively. We observe that for a fixed $\beta_{\rm{edge}}$ and $\alpha_{\rm B}$, the shock radius $r_{\rm{s}}$ moves away from the horizon as $\Upsilon_{\rm s}$ is increased. However, when thermal conduction exceeds its limiting value, the shock disappears as RHCs are not favorable. Furthermore, for a fixed $\Upsilon_{\rm s}$, as $\alpha_{\rm B}$ increases, the shock settles down to a smaller radius due the weakening of the centrifugal repulsion, that results because of the more efficient angular momentum transport. Thereafter, we examine the effect of magnetic field on shock formation. We observe that for gas pressure dominated flow ($i.e.$, $\beta_{\rm edge}=10^4$) with $\alpha_{\rm B}=0.01$, the shock forms further from the black hole at a given $\Upsilon_{\rm s}$. As the magnetic field strength increases ($i.e.$ as $\beta_{\rm edge}$ decreases), the shock radius $r_{\rm s}$ proceeds towards the black hole. The open circles and asterisks joined with solid and dashed lines represent the variation of $r_{\rm s}$ with $\Upsilon_{\rm s}$ for $\beta_{\rm edge}=10^4$ and $\beta_{\rm edge}=500$, respectively. Indeed, the disc's high energy radiation flux is primarily determined by radiative cooling processes, which are strongly dependent on the density $\rho$ and temperature $T$ distributions across the shock front \cite{chakrabarti-titarchuk-1995}~\cite{mandal-chakrabarti-2005}. Accordingly, in Fig. \ref{fig:3}b, we illustrate the variation of the compression ratio $(R)$, which quantifies density compression across the shock, as a function of $\Upsilon_{\rm s}$ for the shock-induced accretion solutions presented in Fig. \ref{fig:3}a. As $\Upsilon_{\rm s}$ increases, the shock generally moves away from the black hole horizon, causing the post-shock region (PSC) to experience less compression and leading to a decrease in the compression ratio $R$. In a way, the shock becomes weaker in the presence of thermal conduction. In contrast, when $\beta_{\rm edge}$ decreases, the shock front moves inward toward the black hole, which results in greater compression of the PSC and, consequently, an increase in $R$. In addition, we also examine the variation of shock strength $S$ as a function of $\Upsilon_{\rm s}$ for the solutions presented in Fig. \ref{fig:3}a and observe that $S$ decreases as thermal conduction is increased, as shown in Fig. \ref{fig:3}c. Based on these finding, it is evident that shock-induced global accretion solutions exist across a wide range of $\Upsilon_{\rm s}$ for various values of $\beta_{\rm edge}$ and $\alpha_{\rm B}$. Notably, these shock-driven accretion solutions have been successful in explaining the observed spectro-temporal characteristics of black hole X-ray binary sources, as demonstrated in numerous studies \cite{chakrabarti-titarchuk-1995, chakrabarti-manickam-2000, mandal-chakrabarti-2005, nandi-etal-2012, iyer-etal-2015, das-etal-2021, majumder-etal-2022, nandi-etal-2024}.

\begin{figure}
\begin{center}
    \includegraphics[width=\columnwidth]{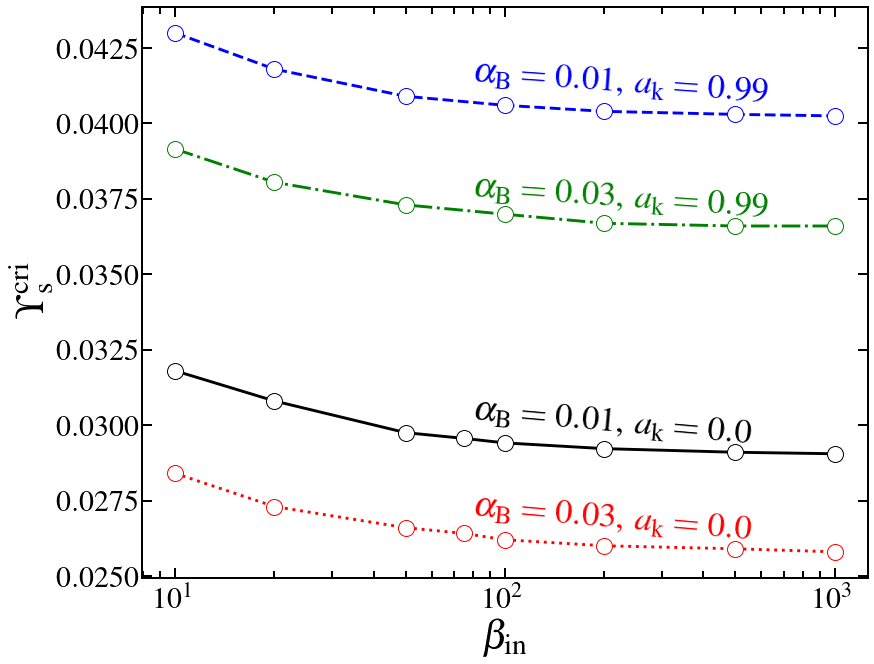}
\end{center} 
\caption{Plot of $\Upsilon_{\rm s}^{\rm{cri}}$ with the $\beta_{\rm{in}}$ for various values of $\alpha_{\rm{B}}$ for non-rotating ($a_{\rm{k}}=0.0$) and maximally rotating ($a_{\rm{k}}=0.99$)  BHs. Here, $\Upsilon_{\rm s}^{\rm{cri}}$ is the limiting value of the $\Upsilon_{\rm s}$ beyond which shock solutions cease to exist; see text for details.
}
\label{fig:4} 
\end{figure}

It is evident that global shock-induced accretion solutions exist within a specific range of the conduction parameter $\Upsilon_{\rm s}$ which is bounded by its critical value $\Upsilon^{\rm cri}_{\rm s}$. Notably, $\Upsilon^{\rm cri}_{\rm s}$ does not have a universal value; instead, it depends on the other model parameters. To explore this, we calculate $\Upsilon^{\rm cri}_{\rm s}$ for both weakly rotating ($a_{\rm k} \rightarrow 0$) and rapidly rotating ($a_{\rm k} \rightarrow 1$)black holes, and investigate how it varies with the magnetic field strength at the inner critical point. Since the inner critical points ($r_{\rm in}$) are located close to the horizon, it is reasonable to assume that the flow enters into the black hole with magnetic fields ($\beta$) similar to those at the inner critical point $\beta_{\rm in}$. Hence, we examine the variation of $\Upsilon^{\rm cri}_{\rm s}$ with $\beta_{\rm in}$ for different $\alpha_{\rm B}$ values and depict the obtained results in Fig. \ref{fig:4}. It is important to note that while calculating $\Upsilon^{\rm cri}_{\rm s}$, we freely vary the remaining model parameters. We observe that $\Upsilon^{\rm cri}_{\rm s}$ increases as the magnetic field strength increaes ($i.e.$ as $\beta_{\rm in}$ decreases) regardless of the black hole spin ($a_{\rm k}$). In addition, higher viscosity leads to lower values of $\Upsilon^{\rm cri}_{\rm s}$. Furthermore, we notice that for a given set of ($\beta_{\rm in}, \alpha_{\rm B}$), $\Upsilon^{\rm cri}_{\rm s}$ is larger for higher black hole spin $a_{\rm k}$ and smaller for lower $a_{\rm k}$.

\begin{figure}
\begin{center}
    \includegraphics[width=\columnwidth]{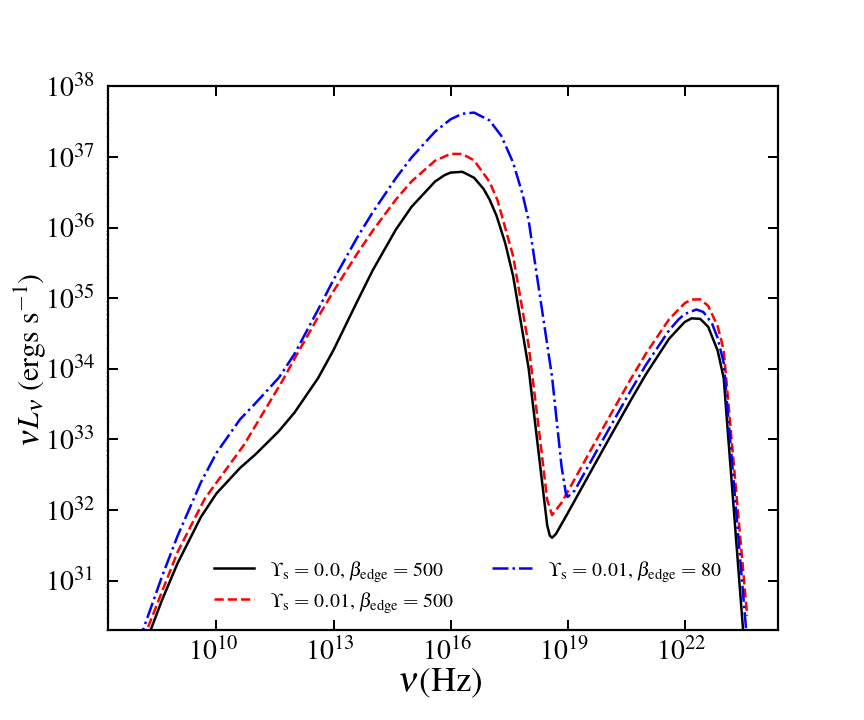}
\end{center}
\caption{\label{fig:5} Spectral energy distribution (SED) of accretion solutions for different conduction parameter ($\Upsilon_{\rm s}$) and plasma-$\beta$. Solid (black), dashed (red) and dot-dashed curves denote results for ($\Upsilon_{\rm s},\beta_{\rm edge}$) $=(0.0, 500)$, $(0.01,500)$ and $(0.01, 80)$, respectively. See the text for details.
}
\end{figure}

Furthermore, we put efforts to calculate the monochromatic  luminosity. For a convergent shocked accretion flow, we obtain, 
\begin{equation}
    \label{eqn:SED}
    L_{\nu} = 2 \int_{r_{\rm edge}}^{r_{\rm h}} \int_{0}^{2\pi}Q^{-}_{\nu}  H r \thickspace dr d\phi \thickspace \thickspace{\rm erg} \thickspace {\rm s}^{-1} {\rm Hz}^{-1},
\end{equation}
where $Q^{-}_\nu$ refers the total emissivity from  at a emission frequency $\nu$. We calculate total emissivity by combining both bremsstrahlung and synchrotron emissivities \cite{novikov-thorne-1973, rybicki-lightman-1986, wardzinski-zdziarski-2001} as $Q^{-}_\nu = Q_{\nu}^{\rm syn} + Q_{\nu}^{\rm brem}$, which are given by,
\begin{align*}
    Q_{\nu}^{\rm brem}&=  6.8 \times 10^{-38} n_{e}^2 T_{e}^{-1/2} \\
    & \times (1+4.4\times10^{-10}T_{e})\exp\left(\frac{h\nu}{k_{B} T_{e}}\right) \quad {\rm and} \\
    Q_{\nu}^{\rm syn} &= \frac{2^{1/6} \pi ^{3/2} e^{2} n_{e} \nu}{3^{5/6} c K_{2}(1/\Theta) u^{1/6}}\exp\left[-\left(\frac{9 u}{2}\right)^{1/3}\right],
\end{align*}
where $u=2 \pi m_{e} c\nu/(e B \Theta^{2})$, and $K_{2}(1/\Theta)$ is modified Bessel function of order two.

In this work, we consider strong coupling between electrons and ions, that results in single temperature accretion flow. However, in a realistic scenario, since electrons are much lighter than ions, the electron temperature ($T_e$) must be lower than the ion temperature ($T_i = T$), at least near the vicinity of the black hole. To account this, we follow the work of \cite[]{Chattopadhyay-Chakrabarti2000} and estimate the electron temperature as $T_ e = \sqrt{(m_e/m_i)} T$, where $m_i$ and $m_e$ are the masses of ions and electrons, respectively. Using Eq. (\ref{eqn:SED}), we calculate the spectral energy distribution (SED) for three different shock-induced accretion solutions with varying $\Upsilon_{\rm s}$ and $\beta_{\rm edge}$ values. The results are shown in Fig. \ref{fig:5}, where the variation of $\nu L_{\nu}$ with frequency $\nu$ is plotted. Solid (black), dashed (red), and dot-dashed (blue) curves represent the results corresponds to $(\Upsilon_{\rm s},\beta_{\rm edge}) = (0.0, 500)$, $(0.01, 500)$, and $(0.01, 80)$, respectively. We observe that for all cases, synchrotron photons dominate the lower energy part of the spectrum, peaking around $\nu \sim 10^{16}$ Hz, while bremsstrahlung photons contribute to the high-energy part, peaking at $\nu \sim 10^{22}$ Hz. The spectra exhibit a sharp cutoff at $ \nu \approx 10^{23}$ Hz, corresponding to an electron temperature $T_e \approx 10^{10}$ K at the inner edge ($r \sim r_{\rm h}$) of the {\red disc}. We find that the peaks of the SEDs are largely insensitive to effect of the thermal conduction ($\Upsilon_{\rm s}$); however, the synchrotron peaks shift to higher frequencies as the  becomes more magnetized ($i.e.$, smaller $\beta_{\rm edge}$). We also observe that the SED is influenced by thermal conduction ($\Upsilon_{\rm s}$), which eventually enhances the disc luminosity. Interestingly, as the magnetic activity increases within the disc, the hot accreting plasma produces more luminous power spectra compared to a weakly magnetized disc. It's worth noting that, as our model framework is developed for low accretion rates, the resulting SED is potentially suited for radiatively inefficient accretion flows, including LLAGNs.

\section{Conclusion}

In this study, we examine the impact of thermal conduction and plasma-$\beta$ on global, transonic, magnetized, viscous, advective accretion flows around rotating black holes in presence of bremsstrahlung and synchrotron cooling processes. In this formalism, accretion disc is threaded by the toroidal magnetic fields and the spacetime geometry is mimicked by the effective potential introduced by \cite[]{dihingia-etal-2018}. Furthermore, we adopt relativistic equation of state (REoS) to describe the thermodynamical flow variables. With this, we solve the governing equations that describe the flow motion in accretion disc and obtain the shock-induced global transonic accretion for set of model parameters, namely accretion rate ($\dot{m}$), viscosity parameter ($\alpha_{\rm{B}}$), black hole spin ($a_{\rm k}$), conduction parameter ($\Upsilon_{\rm s}$), plasma-$\beta$, energy and angular momentum of the flow. Our results establish that magnetic fields and thermal conduction play crucial role in regulating the shock phenomena, influencing shock location ($r_{\rm{s}}$), compression ratio ($R$), and shock strength ($S$), which in turn alters the emission spectrum of the disc. The key findings of this study are outlined as follows:

\begin{itemize}
    \item We find that global transonic magnetized accretion flows undergo shock transitions when thermal conduction is active within the disc. This shock triggering naturally leads to the formation of a hot and dense post-shock flow, which resembles a post-shock corona (PSC) (see Fig. \ref{fig:1}). At the PSC, soft photons from the pre-shock flow can be reprocessed, resulting in the production of hard X-rays, which are commonly observed from black hole sources \cite{chakrabarti-titarchuk-1995, mandal-chakrabarti-2005}.

    \item We observe that both thermal conduction and magnetic fields play a pivotal role in shock formation. In particular, thermal conduction exerts an effect opposite to that of magnetic fields in determining the shock transition (see Fig. \ref{fig:2}). In addition, we find that shocks continue to form in the accretion flow across a wide range of conduction parameter $\Upsilon_{\rm s}$ and plasma-$\beta$, including the viscosity parameter $\alpha_{\rm B}$. In a way, shock transitions are driven by the synergistic interplay between thermal conduction, viscosity, and magnetic fields. These factors collectively influence the shock properties and determine the disc structure of the accretion flow. Moreover, we notice that for all cases, strong shocks are formed when $\Upsilon_{\rm s}$ is small, with shock strength diminishing as $\Upsilon_{\rm s}$ increases (see Fig. \ref{fig:3}).

    \item We calculate the critical conduction parameter $\Upsilon^{\rm cri}_{\rm s}$ that renders global shocked magnetized accretion solutions around both weakly rotating ($a_{\rm k}\rightarrow 0$) as well as rapidly rotating ($a_{\rm k}\rightarrow1$) black holes. Our results reveal that accretion flows around rapidly rotating black holes can sustain higher values of $\Upsilon^{\rm cri}_{\rm s}$ compared to weakly rotating black holes, independent of magnetic field strengths in the inner disc region ($i.e.$, plasma-$\beta$). Furthermore, for a fixed $a_{\rm k}$ and $\beta_{\rm in}$, we find that $\Upsilon^{\rm cri}_{\rm s}$ is larger for weakly viscous flows and decreases with increasing viscosity (see Fig. \ref{fig:4}).

    \item We examine the impact of the conduction parameter $\Upsilon_{\rm s}$ and plasma-$\beta$ on the disc emission spectrum (SEDs) resulted due to the bremsstrahlung and synchrotron cooling processes. Our findings indicate that the inclusion of thermal conduction noticeably enhances the emission spectrum. Moreover, we observe that increased magnetic activity also leads to more luminous emission spectrum (see Fig. \ref{fig:5}).    
\end{itemize}

It is worth mentioning that most earlier studies involving thermal conduction in accretion flows adopted the self-similar approach \cite[and references therein]{tanaka-menou-2006,shadmehri-2008,Faghei2012,Faghei2013,ghoreyshi-shadmehri-2020}. Although self-similar solutions provide useful physical insights, they are inherently limited in describing the global structure of the flow, especially near the inner and outer boundaries of the accretion disc \cite[]{narayan-etal-1997,chen-etal-1997}. Addressing these limitations, the present study is the first, to the best of our knowledge, to explore shock-induced global magnetized accretion flows around a rotating black hole while incorporating the effects of thermal conduction.

Finally, we wish to emphasize that the present formalism is developed under several simplifying assumptions. We approximate the spacetime geometry adopting an effective pseudo-potential instead of a full general relativistic treatment and therefore do not explicitly capture the frame-dragging effect that compels the accretion flow to corotate with the black hole near the event horizon. We focus exclusively on Rankine-Hugoniot shocks, setting aside isothermal and isentropic shocks, although they are also likely to form in accretion flows \cite[]{Abramowicz-Chakrabarti1990,Le-Becker2004,Le-Becker2005,Das-Choi2009,Das-etal2009}. We ignore energy dissipation across the shock front and assume a uniform $\Upsilon_S$ across both upstream and downstream flows. We further consider the toroidal component of the magnetic field, neglecting the poloidal components and the effects of anisotropic thermal conduction in complex magnetic field configurations. Moreover, we neglect mass loss from the disc, even though thermal conduction may play a significant role in driving outflows and/or winds. While all these processes are relevant in accretion dynamics, their inclusion is beyond the scope of this paper and will be explored in future studies.

\section*{Data Availability}

The data underlying this paper will be available with reasonable request.

\section*{Acknowledgments}

Authors thank the anonymous reviewer for constructive comments and useful suggestions that help to improve the manuscript. Authors also thank the Department of Physics, IIT Guwahati, India for providing the infrastructural support to carry out this work.


\appendix

\section{Calculation of wind equation}

After some algebraic manipulation, the equations for radial momentum, azimuthal momentum, and entropy generation are expressed as follows:
\begin{equation}
R_{0} + R_{\Theta}\frac{d \Theta}{d r} + R_{\lambda} \frac{d \lambda}{d r} + R_{\beta} \frac{d \beta}{d r} + R_{\upsilon} \frac{d \upsilon}{d r} = 0, 
\end{equation}
\begin{equation}
L_{0} + L_{\Theta}\frac{d \Theta}{d r} + L_{\lambda} \frac{d \lambda}{d r} + L_{\beta} \frac{d \beta}{d r} + L_{\upsilon} \frac{d \upsilon}{d r} = 0,
\end{equation}
\begin{equation}
B_{0} + B_{\Theta}\frac{d \Theta}{d r} + B_{\lambda} \frac{d \lambda}{d r} + B_{\beta} \frac{d \beta}{d r} + B_{\upsilon} \frac{d \upsilon}{d r} = 0,
\end{equation}
\begin{equation}
E_{0} + E_{\Theta}\frac{d \Theta}{d r} + E_{\lambda} \frac{d \lambda}{d r} + E_{\beta} \frac{d \beta}{d r} + E_{\upsilon} \frac{d \upsilon}{d r} = 0.
\end{equation}

Simplifying the above equations, we have, 

\begin{equation}
\frac{d\upsilon}{d r} = \frac{\mathcal{N}(r,\upsilon,\lambda,\Theta)}{\mathcal{D}(r,\upsilon,\lambda,\Theta)}
\end{equation}

\begin{equation}
\frac{d\Theta}{d r} = \Theta_{\textrm{1}} + \Theta_{\textrm{2}}\frac{d\upsilon}{d r}
\end{equation}

\begin{equation}
\frac{d\lambda}{d r} = \lambda_{\textrm{1}} + \lambda_{\textrm{2}}\frac{d\upsilon}{d r}
\end{equation}

\begin{equation}
\frac{d\beta}{d r} = \beta_{\textrm{1}} + \beta_{\textrm{2}}\frac{d\upsilon}{d r},
\end{equation}
where,
\begin{align*}
& \mathcal{N}(r,\upsilon,\lambda,\Theta)  = -E_{\beta} R_{\lambda} B_{\Theta} B_{0} + E_{\beta} R_{\Theta} B_{\lambda} L_{0} \\
& + E_{0} R_{\lambda} B_{\Theta} L_{\beta} - E_{0} R_{\Theta} B_{\lambda} B_{\beta} + E_{\beta} R_{\lambda} B_{0} L_{\Theta} \\
& - E_{0} R_{\lambda} B_{\beta} L_{\Theta} - E_{\beta} R_{0} B_{\lambda} L_{\Theta} + E_{0} R_{\beta} B_{\lambda} L_{\Theta} \\
& + E_{\lambda} [ R_{\beta} B_{\Theta} L_{0} - R_{0} B_{\Theta} L_{\beta} + R_{\Theta} (-B_{\beta} L_{0} + B_{0} L_{\beta}) \\
& - R_{\beta} B_{0} L_{\Theta} +  R_{0} B_{\beta} L_{\Theta}] - E_{\beta} R_{0} B_{0} L_{\lambda} + E_{0} R_{\Theta} B_{\beta} L_{\lambda} \\
& + E_{\beta} R_{0} B_{\Theta} B_{\lambda} - E_{0} R_{\beta} B_{\Theta} L_{\lambda} + E_{\Theta} [ R_{\lambda} B_{\beta} L_{0} \\
&- R_{\beta} B_{\lambda} L_{0} - R_{\lambda} B_{0} L_{\beta} + R_{0} B_{\lambda} L_{\beta} + R_{\beta} B_{0} L_{\lambda}\\
&- R_{0} B_{\beta} L_{\lambda} ],
& \mathcal{D} (r,\upsilon,\lambda,\Theta) = E_{\beta} R_{\lambda} B_{\theta}L_{\upsilon} - E_{\beta}R_{\Theta}B_{\lambda}L_{\upsilon} \\
& - E_{\upsilon}R_{\lambda}B_{\Theta}L_{\beta} + E_{\upsilon}R_{\Theta}B_{\lambda}L_{\beta} - E_{\beta}R_{\lambda}B_{\upsilon}L_{\Theta} \\
& + E_{\upsilon}R_{\lambda}B_{\beta}L_{\Theta} + E_{\beta}R_{\upsilon}B_{\lambda}L_{\Theta} - E_{\upsilon}R_{\beta} B_{\lambda} L_{\Theta} \\
& + E_{\lambda} [ -R_{\beta}B_{\Theta} L_{\upsilon} +R_{\upsilon} B_{\Theta} L_{\beta} +R_{\Theta}( B_{\beta} L_{\upsilon} \\
& -B_{\upsilon} L_{\beta} ) + R_{\beta} B_{\upsilon}L_{\Theta}  -R_{\upsilon} B_{\beta} L_{\Theta} ]  + E_{\beta} R_{\Theta} B_{\upsilon} L_{\lambda} \\
& - E_{\upsilon} R_{\Theta} B_{\beta} L_{\lambda} - E_{\beta} R_{\upsilon} B_{\Theta} L_{\lambda} + E_{\upsilon} R_{\beta} B_{\Theta}L_{\lambda} \\
& + E_{\Theta} [ - R_{\lambda} B_{\beta} L_{\upsilon} + R_{\beta} B_{\lambda} L_{\upsilon} + R_{\lambda} B_{\upsilon} L_{\beta} \\
& - R_{\upsilon} B_{\lambda} L_{\beta} - R_{\beta} B_{\upsilon} L_{\lambda} + R_{\upsilon} B_{\beta} L_{\lambda} ]\\
\Theta_{1} & = \frac{\Theta_{11}}{\Theta_{33}}, \thickspace \Theta_{2} = \frac{\Theta_{22}}{\Theta_{33}}, \thickspace \lambda_{1} = \frac{\lambda_{11}}{\lambda_{33}}, \thickspace \lambda_{2} = \frac{\lambda_{22}}{\lambda_{33}},\\
\beta_{1} & = \frac{\beta_{11}}{\beta_{33}}, \thickspace \beta_{2} = \frac{\beta_{22}}{\beta_{33}}, \\
\Theta_{11} &= -  [(E_{\beta}L_{0} - E_{0} L_{\beta} )( E_{\lambda} B_{\beta} - E_{\beta} B_{\lambda} )  \\
& + (-E_{\lambda} L_{\beta} + E_{\beta} L_{\lambda}) ( E_{\beta}B_{0} -E_{0} B_{\beta} )],\\
\Theta_{22} & = -[ (E_{\beta} L_{\upsilon} - E_{\upsilon} L_{\beta}) ( E_{\lambda} B_{\beta} - E_{\beta} B_{\lambda} ) \\
& + ( E_{\beta} \beta_{\upsilon} - E_{\upsilon} B_{\beta} ) ( -E_{\lambda} L_{\beta} + E_{\beta} L_{\lambda} ) ],\\
\Theta_{33} & = (E_{\lambda}B_{\beta} -E_{\beta} B_{\lambda} ) ( -E_{\Theta} L_{\beta} + E_{\beta} L_{\Theta} ) \\
& + ( -E_{\Theta} B_{\beta} + E_{\beta} \beta_{\Theta} ) ( -E_{\lambda} L_{\beta} + E_{\beta} L_{\lambda} ),\\
\lambda_{11} & =  -E_{\Theta} B_{\beta} L_{0} + E_{\beta} B_{\Theta} L_{0} + E_{\Theta} B_{0} L_{\beta} - E_{0} B_{\Theta} L_{\beta}, \\
& - E_{\beta} B_{0} L_{\Theta} + E_{0} B_{\beta} L_{\Theta}, \\
\lambda_{22} & =  -E_{\Theta} B_{\beta} L_{\upsilon} + E_{\beta} B_{\Theta} L_{\upsilon} + E_{\Theta} B_{\upsilon} L_{\beta} - E_{\upsilon} B_{\Theta} L_{\beta}, \\
& - E_{\beta} B_{\upsilon} L_{\Theta} +E_{0} B_{\beta} L_{\Theta}, \\
\lambda_{33} & = E_{\lambda} B_{\Theta} L_{\beta} - E_{\Theta} B_{\lambda} L_{\beta} - E_{\lambda} B_{\beta} L_{\Theta} + E_{\beta} B_{\lambda} L_{\Theta} \\
& + E_{\Theta} B_{\beta} L_{\lambda} - E_{\beta} B_{\Theta} L_{\lambda},\\
\beta_{11} & =  - E_{\lambda} B_{\Theta} L_{0} + E_{\Theta} B_{\lambda} L_{0} + E_{\lambda} B_{0} L_{\Theta} - E_{0} B_{\lambda} L_{\Theta} \\
& - E_{\Theta} B_{0} L_{\lambda}+ E_{0} B_{\Theta} L_{\lambda},\\
\beta_{22} & = -E_{\lambda}B_{\Theta}L_{\upsilon} + E_{\Theta} B_{\lambda}L_{\upsilon} + E_{\lambda} B_{\upsilon}L_{\Theta} - E_{\upsilon} B_{\lambda}L_{\Theta} \\
& - E_{\Theta} B_{\upsilon}L_{\lambda} + E_{\upsilon} B_{\Theta}L_{\lambda},\\
\beta_{33} & = \lambda_{33},\\
R_{0} & =  \frac{4\Theta}{r \beta \tau} + \frac{2 \Theta (1+\beta^{-1})}{\tau h} \left( -\frac{3}{2 r} + \frac{F_{3}}{2 F_{2}} - \frac{1}{2\Delta}\frac{d \Delta}{d r} \right)\\
& + \frac{d \Psi_{\rm{e}}^{\rm{eff}}}{d r}, R_{\upsilon} = \upsilon - \frac{2\Theta}{\tau h \upsilon}(1+\beta^{-1}), \\
R_{\Theta} & = \frac{(1+\beta^{-1})}{\tau h}, R_{\lambda} = \frac{F_{4} \Theta (1+\beta^{-1})}{\tau h F_{2}}, R_{\beta} = - \frac{\Theta}{\tau h \beta^2}, \\
L_{0} & = -2 \alpha_{\rm{B}} \upsilon^2 -\frac{4 \alpha_{\rm{B}} \Theta (1+\beta^{-1}) }{\tau} \\
& +  \frac{d \Delta}{dr}\left( \frac{r \alpha_{\rm{B}} (\tau \upsilon^2 \beta + 2 \Theta(1+\beta^{-1}) )}{2 \tau \beta} \right),\\
L_{\upsilon} &= -r \alpha_{\rm{B}} \upsilon + \frac{2 r \alpha_{\rm{B}} \Theta (1+\beta^{-1}) }{\tau \upsilon},\\
L_{\Theta} & = - \frac{2 r \alpha_{\rm{B}} (1+\beta^{-1}) }{\tau}, \\
L_{\lambda} & = \upsilon, \thickspace L_{\beta} = \frac{2 r \alpha \Theta}{\tau \beta^2},\\
\end{align*}
\begin{align*}
E_{0} & = -\frac{Q^{-}}{\rho H} +  \frac{5 \Upsilon_{\rm s} \Theta }{\tau}\sqrt{\frac{2 \Theta}{\tau}} \left( \frac{1}{r} -\frac{F_{3}}{F_{2}} + \frac{1}{\Delta}\frac{d \Delta}{d r} \right) \\
& + \frac{3 \upsilon \Theta}{r \tau} - \frac{F_{3}\upsilon \Theta}{\tau F_{2}} -\frac{2 r \alpha_{\rm{B}} \Theta \omega_{1} (1+\beta^{-1})}{\tau} \\
& - r\alpha_{\rm{B}} \upsilon^2 \omega_{1} +\frac{\upsilon \Theta}{\tau \Delta}\frac{d \Delta}{d r}, \\
E_{\upsilon} &= \frac{2\Theta}{\tau}+\frac{10\sqrt{2}\Upsilon_{\rm s}}{\upsilon}\left( \frac{\Theta}{\tau} \right)^{3/2}, \\
E_{\Theta} & = -\frac{5 \Upsilon_{\rm s}}{\Theta} \left(\frac{2 \Theta}{\tau}\right)^{3/2}  + \frac{\left(1+ 2 N\right) \upsilon}{\tau},\\
E_{\lambda} & = - \frac{F_{4} \upsilon \Theta}{\tau F_{2}} - \frac{5\sqrt{2} \Upsilon_{\rm s} F_{4}}{F_{2}}\left( \frac{\Theta}{\tau} \right)^{3/2} \\
& - r \alpha_{\rm{B}} \left( \upsilon^2 +\frac{2 \Theta(1+\beta^{-1})}{\tau} \right)\omega_{2},\\
    & E_{\beta} = -\frac{1}{\beta(1+\beta)}\left( \frac{\upsilon\Theta}{\tau} + \frac{5\Upsilon_{\rm s} \Theta}{\tau} \sqrt{\frac{2 \Theta}{\tau}}\right),\\
    &B_{0} = \frac{3}{4 r}+ \frac{\zeta}{r}-\frac{F_{3}}{4 F_{2}}-\frac{1}{4 \Delta}\frac{d \Delta}{d r}, \thickspace B_{\upsilon} = \frac{1}{2\upsilon}, B_{\Theta} = \frac{3}{4 \Theta},\\
    & B_{\lambda} = -\frac{F_{4}}{4F_{2}}, \thickspace B_{\beta}= -\frac{1}{(1+\beta)}\left( \frac{1}{2} + \frac{3}{4\beta}\right),\\
    & F_{3} = \frac{F_{1} \lambda \omega_{1}}{(1-\lambda\Omega)^{2}} +  \frac{1}{1-\lambda\Omega} \frac{d F_{1}}{d r},\\
    & F_{4} = \frac{F_{1} \Omega}{(1-\lambda\Omega)^{2}} + \frac{F_{1} \lambda \omega_{2}}{(1-\lambda\Omega)^{2}},\\
    & F_{2} = \frac{1}{(1-\lambda\Omega)} F_{1} ,\\
    & F_{1} = \frac{\left((r^{2}+a_{\textrm{k}}^{2})^{2} + 2 \Delta a_{\textrm{k}}^2\right)}{\left((r^{2}+a_{\textrm{k}}^{2})^{2} - 2 \Delta a_{\textrm{k}}^2\right)}, \\
    & \frac{d F_{2}}{dr} = F_{3} + F_{4}\frac{d \lambda}{d r},\thickspace \frac{d \Omega}{d r} = \omega_{1}+\omega_{2}\frac{d \lambda}{d r},\\
    & \omega_{1} = -\frac{2\left(a_{\textrm{k}}^{3} + 3 a_{\textrm{k}} r^{2}+ \lambda(a_{\textrm{k}} \lambda - 2 a_{\textrm{k}}^{2} + r^{2}(r-3))\right)}{\left(r^{3} + a_{\textrm{k}}^2 (r+2) - 2 a_{\textrm{k}} \lambda\right)^{2}},\\
    & \omega_{2} = \frac{r^{2}\left(a_{\textrm{k}}^2 + r (r - 2)\right)}{\left(r^{3} + a_{\textrm{k}}^2 (r+2) - 2 a_{\textrm{k}} \lambda\right)^{2}}.
\end{align*}

\end{document}